\documentclass[preprint,showpacs,preprintnumbers,amsmath,amssymb]{revtex4}
\usepackage{graphicx}
\usepackage{dcolumn}
\usepackage{bm}

\begin{document}

\title
{\bf Excitonic condensation under spin-orbit coupling and BEC-BCS crossover}

\author{Tu\v{g}rul Hakio\u{g}lu$^{(1)}$ and Mehmet \c{S}ahin$^{(2)}$}

\affiliation{ ${\bf (1)}$ {Department of Physics and National
Nanotechnology Research Center, Bilkent University, 06800 Ankara,
Turkey} \break ${\bf (2)}$ {Department of Physics, Faculty of
Sciences and Arts, Sel\c{c}uk University, Kampus 42075 Konya,
Turkey}}

\begin{abstract}
The condensation of electron-hole (e-h) pairs is studied at zero
temperature and in the presence of a weak spin-orbit coupling (SOC)
in the inversion-layer quantum wells. Under realistic conditions, a
perturbative SOC can have observable effects in the order parameter
of the experimentally long-searched-for excitonic condensate.
Firstly, the fermion exchange symmetry is absent for the e-h pairs
indicating a counterexample to the known classification schemes of
fermion pairing. With the lack of fermion exchange, the condensate
spin has no definite parity. Additionally, the excitonic SOC breaks
the rotational symmetry yielding a complex order parameter in an
unconventional way, i.e. the phase pattern of the order parameter is
a function of the condensate density. This is manifested through
finite off diagonal components of the static spin susceptibility,
suggesting a new experimental method to confirm an excitonic
condensate.
\end{abstract}


\maketitle A rich variety of low temperature collective phases had
been proposed for semiconductors in the 1960s. Condensation of the
e-h pairs was studied primarily by Moskalenko\cite{SAM}, Blatt et
al.\cite{BBB} and the group led by Keldysh \cite{LK}. As the
excitonic density is varied, these phases range from the low density
excitonic BEC to a BCS type ground state at higher densities and
eventually to the e-h liquid\cite{CN,CG,Mosk}.  Initially, the
experimental progress was slow given the difficulties in producing
sufficiently long-lived exciton pairs at low temperatures. One of
the earliest experiments was carried out by Snoke et al.\cite{exp1}
and Hara et al\cite{exp2} on $Cu_2 O$ on 3D samples. The
difficulties were overcome by utilizing indirect excitonic
transitions\cite{Butov}. Still longer lifetimes were obtained by
containing the two dimensional (e) and (h) gases (2DEG and 2DHG)
separately in Coulomb coupled QWs with a stabilizing
E-field\cite{Fukuzawa}. Currently, coupled QWs with improved
lifetimes in the microsecond range, provide optimum experimental
conditions for observing this long proposed state\cite{recentexp}.

Here we investigate exciton condensation (EC) in inversion layer
coupled quantum wells (QW) in the presence of a weak in-plane Rashba
spin-orbit coupling (SOC)\cite{Rashba}. In such systems, and in
contrast to the conventional pairing between identical
fermions\cite{Gorkov}, the only manifested symmetry is time
reversal. The e-h exchange symmetry is absent and the parity of the
condensate mixes with the condensate spin, disabling the
conventional classification schemes\cite{BW}. In turn, there is no
relation between the parity and the spin of the condensate wave
function. Another crucial difference from identical fermion pairing
is that the hole SOC breaks the underlying symmetry of the electron
SOC -known as $C_{\infty v}$- and the corresponding complex
excitonic order parameter in the up-down spin channel develops an
unconventional phase pattern. The latter can be measured in the off
diagonal components of the static spin susceptibility which may be
crucial as a complementary method for identifying the excitonic
condensate.

\begin{figure}
\includegraphics[scale=1.0]{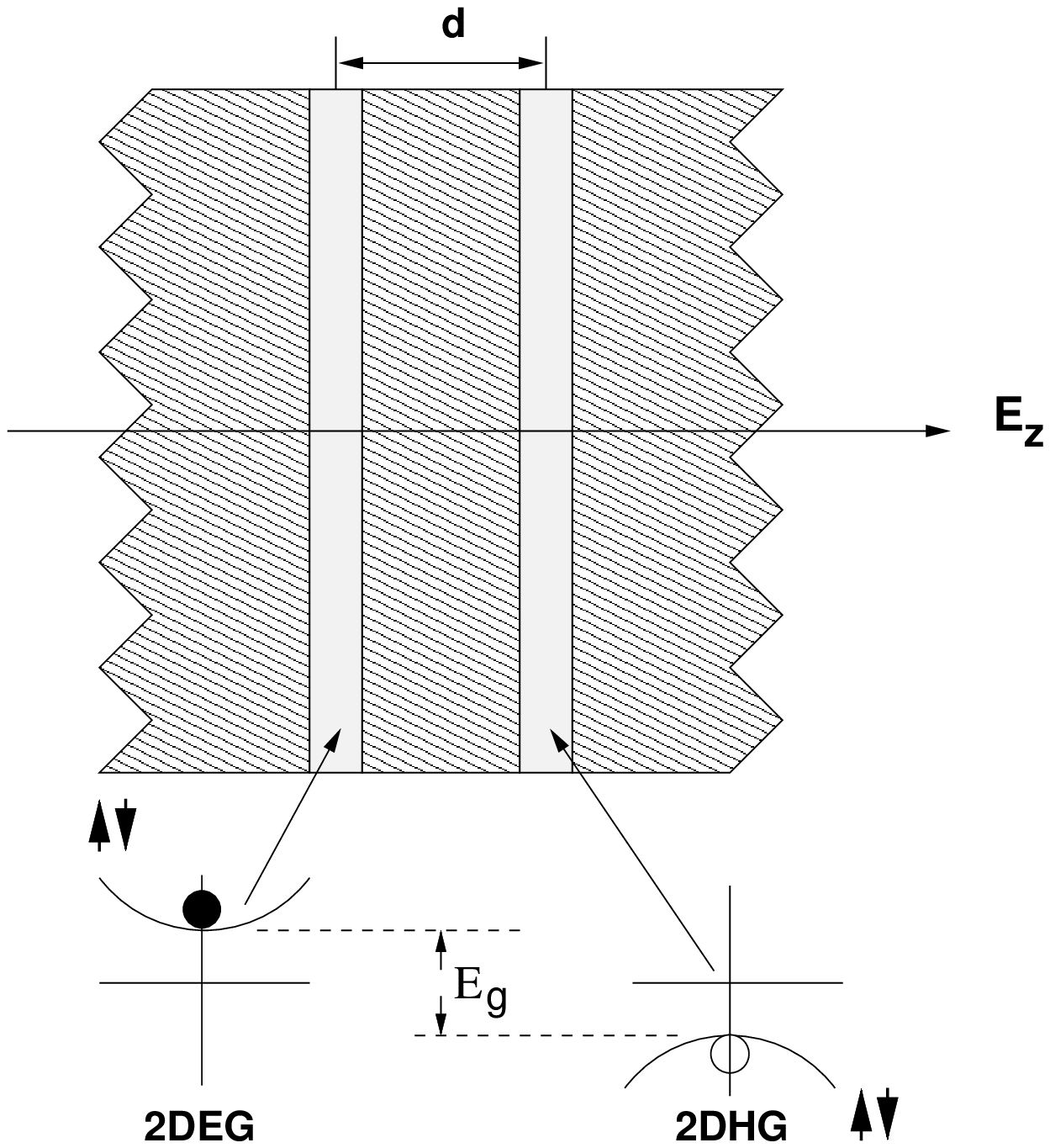}
\caption{The double-well geometry in $x-y$ plane.
The 2DEG and the 2DHG are produced within the GaAs wells inserted in
 high AlGaAs tunneling barriers. We ignore the well widths in this work.
The spin-degenerate conduction and valence subbands are considered
within the parabolic approximation.}
\label{geometry}
\end{figure}

The model geometry studied here is closely related to that of Zhu et
al.\cite{Zhu,PBL1} as illustrated in Fig. \ref{geometry}. The (e)
and the (h) QWs are separated by a high tunneling barrier of
thickness $d$ ($d \simeq 100 \AA$ here). Although typical external
E-fields (for instance Ref.[\cite{Fukuzawa}]) are in the range of
$3-5 kV/cm$, the intrinsic fields due to doping can be as high as
$100-200 kV/cm$, e.g. Ref. [\cite{soexp}]. In this case, it is known
that the Rashba SOC is the dominant mechanism for the splitting of
the energy bands\cite{Bas}. High tunability factors of the SOC by
E-fields was previously shown\cite{nitta,Lu,Grund,Papa} for
inversion layers and the efforts toward much higher tunabilities are
crucial for potential device applications\cite{DattaDas}.

The mechanism of EC is the interband attractive Coulomb interaction.
We consider equal electron and hole densities and the tunneling is
negligible\cite{recentexp}. The intraband Coulomb strengths for a
typical concentration $n_x \simeq  10^{11} cm^{-2}$ are
$V^{ee}=V^{hh}=2\pi e^2/(\epsilon r_{ee})\simeq 4-5 meV$. The layer
separation $d \simeq 1$ in units of the effective Bohr radius
$a_e^*=\hbar^2 \kappa/(e^2 m_e^*) \simeq 100 \AA$. The strength of
the Coulomb interaction between the layers is $V^{eh}=2\pi
e^2/(\epsilon r_{eh}) \simeq 1-2 meV$. The $r_{ee}$ and $r_{eh}$ are
the average e-e (or h-h) and e-h separations. Here the SOC is weak
at typical densities and treated perturbatively in the condensed
excitonic background\cite{footnote1}.

In a typical excitonic semiconductor, the electrons in the
conduction band are in an $s$-like state. For intermediate $n_x$
values it is sufficient to consider the electron-heavy hole (hh)
coupling, with the hh's predominantly in $p$-like
orbitals\cite{Winkler}. The SOCs for the electrons and the hh's are

\begin{equation}
{\cal  H}_{e}=i\alpha_e\,E_z (\sigma_+ k_{-}-\sigma_-k_+)~,
\quad
{\cal  H}_{h}=i\beta_h\,E_z (\sigma_{+} k_{-}^3-\sigma_{-} k_{+}^3)
\end{equation}

where $\sigma_{\pm}=(\sigma_x\pm i \sigma_{y})/2$ are the Pauli
matrices and $k_{\pm}=k_x \pm i k_y$ are the in-plane wavevectors.
The SOC constants $\alpha_e$ and $\beta_h$ can be inferred from many
recent works\cite{soexp,Winkler1,Winkler}. However the agreement on
the suggested values is still lacking.  The values for electrons
vary from $\alpha_e \simeq 30.6 e \AA^2$ to $\alpha_e \simeq 300 e
\AA^2$ in the range from $n_x \le 1\times 10^{11} cm^{-2}$ to $n_x
\simeq 2.2 \times 10^{12} cm^{-2}$. For hh's the only results that
the authors are aware of are by Winkler et al.\cite{Winkler1} in
which $\beta_h=7.5\times 10^{6} e\AA^4$ for $n_x \le 10^{11}
cm^{-2}$. The calculated $\beta_h$ values are however found to be
strongly dependent on the density\cite{Winkler1}. The E-field
strength at the interface generated by the space charges was
estimated by $E_z=e n_x/(2 \epsilon)$. Typical SOC energies for
intermediate $n_x$ covering $10^9 < n_x < 10^{11} (cm^{-2})$ are
perturbatively weaker than typical Coulomb energies at a given $n_x$
as shown in Table I.

\begin{table}
\begin{center}
\caption{Interface E-fields and SOC energies for typical
densities. Here (a)=Ref.[\cite{Winkler}] and (b)=Ref.[\cite{soexp}]}
\begin{tabular}{l|c|c|c|c}
\hline
\hline
$n_x (cm^{-2})$ & $E_z (kV/cm)$ & $\alpha_e k_F E_z (meV)$ &
$\beta_h k_F^3 E_z (meV)$ & $\beta_h k_F^2/\alpha_e$
\\
\hline $10^{9}$ & 1.45 & $5\times 10^{-4}$ & $1.5\times 10^{-4}$ & 0.3 \\
 $10^{10}$ & 14.5 & $1.5\times10^{-2}$ & $4.9\times 10^{-2}$ &  3.3 \\
 $10^{11}$ $^{(a)}$ & 145 & 0.8 & 15 & 18.75 \\
 $10^{12}$ $^{(b)}$ & $1.45 \times 10^{3}$ & 154 & $4.8 \times 10^{3}$ & 31 \\
\hline
\hline
\end{tabular}
\end{center}
\label{table1}
\end{table}

An order of magnitude increase for the $\alpha_e$ values was also
estimated for In-based QWs in Ref[\cite{Bas}]. The electronic $r_s$
values vary in the range $1 \le r_s < 500$. The regime characterized
by $r_s \simeq 2-5$ as a crossover\cite{CN,CG,Zhu} from the strongly
interacting BEC to the weakly interacting BCS type condensation can
therefore be probed by the change in the behavior of the spin-orbit
order parameter.

For $n_x < 10^{11} cm^{-2}$, only the lowest hh states are occupied
in the valence band\cite{Winkler}. For $n_x \le 10^{9} cm^{-2}$ the
spin dependent splitting is difficult to observe\cite{Tutuc}. We
therefore consider here the range $10^{9} \le n_x \le 10^{11}
(cm^{-2})$.

In the absence of SOC the condensed state is formulated by the e-hh
quasi particle eigenstates $\hat{\alpha}_{{\vec
k},\sigma}=\cos(\theta_{\vec k}/2)\hat{c}_{{\vec k},\sigma}-
\sin(\theta_{\vec k}/2)\hat{d}^\dagger_{-{\vec k},\sigma}~;~~
\hat{\beta}_{{\vec k},\sigma}=\sin(\theta_{\vec k}/2)\hat{c}_{{\vec
k},\sigma}+ \cos(\theta_{\vec k}/2)\hat{d}^\dagger_{-{\vec
k},\sigma}$ where ${\vec k}=(k_x,k_y)$, $\sigma=\uparrow,
\downarrow$ and $\hat{c}_{{\vec k},\sigma}$ and $\hat{d}_{{\vec
k},\sigma}$ are the annihilation operators for the electron and the
hh. The cosine and sine coherence factors have been found\cite{Zhu}
for the geometry of Fig. \ref{geometry} using the Hartree-Fock
mean-field of the real excitonic order parameter (EOP)

\begin{equation}
\Delta_{0}^{(\sigma \sigma^\prime)}({\vec k})=\sum_{{\vec k}^\prime}
V^{eh}_{{\vec k}-{\vec k}^\prime} \langle \hat{c}^{\dagger}_{{\vec
k}^\prime,\sigma} \hat{d}^\dagger_{-{\vec
k}^\prime,\sigma^\prime}\rangle~. \label{the_gap}
\end{equation}

Due to the rotational invariance of the momentum and the spin spaces
separately, the ground state is isotropic and spin
independent\cite{Halperin-Rice}. At low $n_x$, the EOP is large near
$k=0$ and the condensation is BEC type\cite{Zhu}. For increasing
$n_x$ the peak position shifts to a finite value near $k_F$ where
the BCS type pairing is dominant\cite{PBL1}.

Including the spin-orbit effect, the time reversal symmetry remains but
the spin degeneracy is lifted. The full Hamiltonian, in the basis
$( \hat{\alpha}_{{\vec k},\uparrow}\,
\hat{\beta}_{-{\vec k},\uparrow}^\dagger \, \hat{\alpha}_{{\vec k},\downarrow} \,
\hat{\beta}_{-{\vec k},\downarrow}^\dagger )$ is then


\begin{equation}
\cal H = \left( {\begin{array}{*{20}c}
   { - E_{\vec k} } & 0 & {iA + \Delta _1 } & {iC + \Delta _2 }  \\
   0 & {E_{\vec k} } & { - iC + \Delta _3^* } & {iB - \Delta _1 }  \\
   { - iA^*  + \Delta _1^* } & {iC^*  + \Delta _3 } & { - E_{\vec k} } & 0  \\
   { - iC^*  + \Delta _2^* } & { - iB^*  - \Delta _1^* } & 0 & {E_{\vec k} }  \\
\end{array} } \right)
\label{hamilt.1}
\end{equation}

\noindent where the diagonal terms correspond to the lower
($\hat{\alpha}_{{\vec k},\sigma}$) and the upper
($\hat{\beta}_{{\vec k},\sigma}$) excitonic bands determined
by\cite{Zhu}

\begin{eqnarray}
E_{\vec k}&=&\sqrt{\zeta_{\vec k}^2+\Delta_{0}^2({\vec k})} \nonumber \\
\zeta_{\vec k}&=&E_g/2+\epsilon_{\vec k}-\mu_x+\sum_{{\vec k}^\prime}
V_{{\vec k}-{\vec k}^\prime}\zeta_{{\vec k}^\prime}/E_{{\vec k}^\prime}
=E_{\vec k}\cos\theta_{\vec k} \nonumber \\
\Delta_{0}({\vec k})&=&\frac{1}{2}\sum_{{\vec k}^\prime}
V^{eh}_{{\vec k}-{\vec k}^\prime}\,\Delta_{0}({\vec k}^\prime)/E_{{\vec k}^\prime}
= E_{\vec k}\sin\theta_{\vec k} \label{pbl} \\
n_x&=&\frac{1}{2}
\sum_{\vec k^\prime}(1-\zeta_{{\vec k}^\prime}/E_{{\vec k}^\prime}) \nonumber
\end{eqnarray}

\noindent Here $\mu_x$ is the exciton chemical potential. In
(\ref{hamilt.1}) $A$ and $B$ are the intraband excitonic SOCs for
the lower and the upper branches and $C$ is the interband SOC. The
higher excitonic band can be neglected here due to the fact that the
$\hat{\beta}$ states contribute to the $\hat{\alpha}$ state
intraband transition energies on the order of $\vert C
\vert^2/\Delta_0^2$ for low momenta,  and $\vert C
\vert^2/\zeta_{\vec k}^2$ for high momenta, which are both
negligible. Eliminating the $\hat{\beta}$-like states, the
Hamiltonian
 can be reduced to a $2\times 2$ matrix for the lower band where only
$A$ and $\Delta_1$ are relevant which are

\begin{eqnarray}
A({\vec k})&=&
iE_z[\alpha_e\cos^2(\theta_{\vec k}/2)k_{-}+\beta_h\sin^2(\theta_{\vec k}/2)k_{-}^3]
\qquad \label{rashba.couplings} \\
\Delta_1({\vec k})&=&\cos(\theta_{\vec k}/2)\sin(\theta_{\vec k}/2)
[\Delta_{\uparrow \downarrow}^*({\vec k})+
\Delta_{\downarrow \uparrow}({\vec k})] \label{delta_n} \qquad
\label{rashba.order_par}
\end{eqnarray}

\noindent where $\Delta_{\downarrow \uparrow}({\vec k})=\sum_{{\vec
k}^\prime} V^{eh}_{{\vec k}-{\vec k}^\prime} \langle
\hat{c}^{\dagger}_{{\vec k}^\prime,\downarrow}
\hat{d}^\dagger_{-{\vec k}^\prime,\uparrow}\rangle$ is the complex
{\it excitonic spin-orbit order parameter} (ESOOP). The time
reversal dictates that $\Delta_{\downarrow \uparrow}(-{\vec k})=
-\Delta_{\uparrow \downarrow}^*({\vec k})$. The latter also implies
that $\Delta_1({\vec k})$ is odd under ${\vec k} \to -{\vec k}$
although there is no such definite symmetry for the ESOOP. For this
lower branch, the SOC-split eigenenergies are

\begin{equation}
\label{rashba_eigen}
\lambda^{\pm}_{\vec k}=-E_{\vec k} \pm \Delta E_{\vec k}~,\qquad \Delta
E_{\vec k}=\vert i A({\vec k})+\Delta_1({\vec k}) \vert
\end{equation}

\noindent where the eigenstates indexed by $\pm$ are


\begin{equation}
\hat \eta _{\vec k, \pm }  \to \frac{1}{{\sqrt 2 }}\left(
{\begin{array}{*{20}c}
   1  \\
   { \pm e^{i\Lambda _{\vec k} } }  \\
\end{array}} \right)
\label{spin_orbit_basis}
\end{equation}

\noindent in the $(\hat{\alpha}_{{\vec k},\uparrow}~,
\hat{\alpha}_{{\vec k},\downarrow})$ basis, and the relative phase
is

\begin{equation}
e^{i\Lambda_{\vec k}}=[i A({\vec k}) +\Delta_1({\vec k})]/
[\vert i A({\vec k})+\Delta_1({\vec k}) \vert]~.
\label{lambda.1}
\end{equation}

The complex ESOOP is then calculated by  Eq.(\ref{rashba.order_par}) as

\begin{equation}
\Delta_1({\vec k})=\frac{1}{4}\frac{\Delta_{0}({\vec k})}{E_{\vec k}}
\sum_{{\vec k}^\prime}e^{i\Lambda_{{\vec k}^\prime}}
V^{eh}_{{\vec k}-{\vec k}^\prime}
\frac{\Delta_{0}({\vec k}^\prime)}{E_{{\vec k}^\prime}}~.
\label{delta.1}
\end{equation}

\noindent Eqs.(\ref{rashba.couplings}), (\ref{rashba.order_par}),
(\ref{lambda.1}) and (\ref{delta.1}) form a self-consistent set
describing the effect of the SOC and they depend on the solutions of
(\ref{pbl}). For electrons there is a $C_{\infty v}$ symmetry
respected by the electronic part of the Hamiltonian\cite{Gorkov}.
This symmetry arises due to continuous rotations in ${\vec k}$-space
and the double covering of the spin-1/2 representation. On the other
hand, the SOC for the hhs has a cubic momentum dependence in
contrast to the linear one in the electronic SOC. Additionally, the
spin space of the hhs (i.e. $S=3/2, S_z=\pm 3/2$) is incomplete in
the spin-3/2 representation. Therefore, the hole SOC breaks the
electronic $C_{\infty v}$ and this has observable consequences.

\begin{figure}
\includegraphics[scale=0.8]{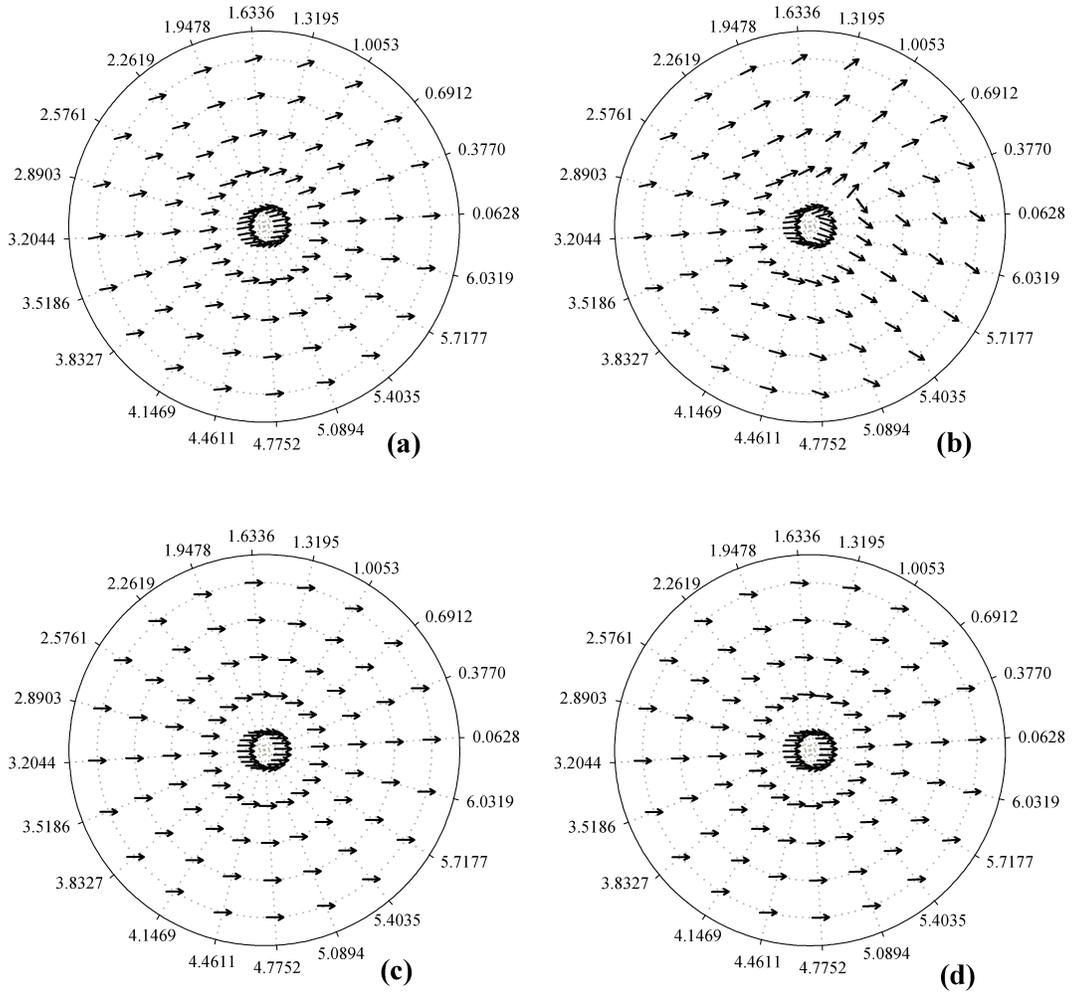}
\caption{Phase of $\Delta_1$ is shown in ${\vec k}$
for $n_x=10^{10} cm^{-2}$ with $E_z=15 kV/cm$ (a) and $E_z=150 kV/cm$ (b);
$n_x=1.77 \times 10^{11} cm^{-2}$ with $E_z=15 kV/cm$ (c), and $E_z=150 kV/cm$ (d).
The radial range is $0\le k\le 3$ in units of $a_e^*\simeq 100 \AA$.}
\label{robust_phase}
\end{figure}

The phase of $\Delta_1$ is plotted in Fig. \ref{robust_phase}. We
observe that at relatively high $n_x$ ($n_x\simeq 1.7 \times 10^{11}
cm^{-2}$ here), the ESOOP phase is relatively coherent for weak electric 
fields, i.e. Fig.\ref{robust_phase} (a), despite the strong variations 
in the SOC. We
attribute this to the dominant contribution in (\ref{delta.1}) near
the Fermi level\cite{Zhu, PBL1} where $\Delta_0(k_F)/E_{k_F} \simeq
1$. There, SOC dictates the phase profile due to a high density of
states (DOS). Thus an increase in $E_z$ has a significant effect as 
observed in Fig. \ref{robust_phase} (b). At lower 
$n_x (\simeq 10^{10} cm^{-2} here)$ there is a weak overlap between the 
condensed pairs and the dominant contribution to (\ref{delta.1}) is near 
$k=0$ where $\Delta_0({\vec k}=0)/E_{{\vec
k}=0} \simeq 1$. The DOS has a minimum there and a small number states 
cannot accomodate the anisotropy in the weak SOC. Thus the phase 
rigidity is imposed by the dominant Coulomb interaction as in 
Fig. \ref{robust_phase} (c). There the phase is less sensitive to the 
E-field strength of the already weak SOC Fig. \ref{robust_phase} (d). 

\begin{figure}
\includegraphics[scale=1.1]{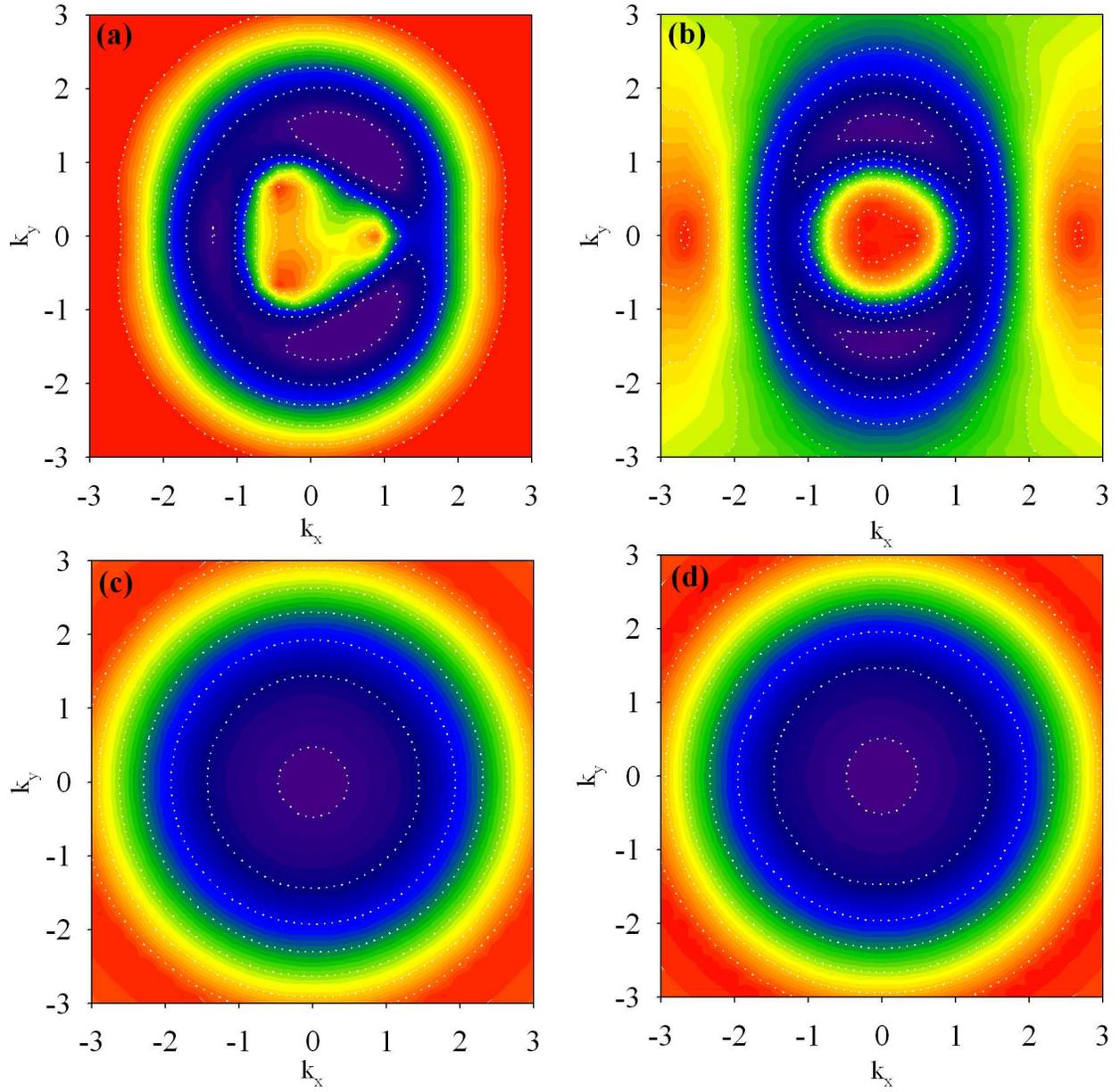}
\caption{Lower excitonic band ($\lambda_{\vec k}^{-}$) is
shown here for the same $E_z$ and $n_x$ values and in the same order
as in Fig. \ref{robust_phase} above. The darker colors mean lower values.}
\label{exc.energy}
\end{figure}

The corresponding solutions for the lower band $\vert \lambda_{\vec
k}^{(-)} \vert$ in Fig. \ref{exc.energy} demonstrate that the
rotational symmetry of the ground state energy is broken by the
anisotropic phase of the SOC. This should be compared with the
isotropic results previously calculated\cite{Zhu,PBL1} without the
SOC. The difference is made by $\Delta E_{\vec k}$ in
(\ref{rashba_eigen}) and it is an interference effect as shown
below. From Fig. \ref{robust_phase} we know that for high $n_x$,
$\Delta_1 $ is phase coherent whereas $A$ is very anisotropic and
complex. Hence an interference is observed in $\vert iA
+\Delta_1\vert$ between these two terms [Fig. \ref{exc.energy} (a)
and (b)]. In the opposite limit of low $n_x$ as shown in Fig.
\ref{exc.energy} (c) and (d), the phase of $\Delta_1$ is imposed by
the SOC [shown by Fig. \ref{robust_phase} (b) and (d)], the
interference is weak  and the energy profile is nearly isotropic.

Other features of Fig. \ref{exc.energy} are similar to the case
without the SOC. At higher $n_x$ the spin independent EOP has a
maximum\cite{Zhu,PBL1} and $\vert \lambda_{\vec k} \vert$ develops a
minimum in the vicinity of the $k_{min} \simeq 1$ ring created by
the pure excitonic term in (\ref{rashba_eigen}), i.e. a BCS type
pairing. In the presence of the SOC, this ring shaped minimum is
deformed as shown in Fig. \ref{exc.energy} (a) and (b). For lower
$n_x$, as shown in Fig. \ref{exc.energy} (c) and (d), the spin
independent EOP is maximum and $\vert \lambda_{\vec k} \vert$ is
minimum at $k_{min}=0$, i.e. a BEC type pairing.  With the SOC, the
additional splitting given by $\vert iA+\Delta_1\vert$ is also
isotropic and does not deform the isotropic contribution of the
spin-independent part.

From the experimental point of view, the off diagonal components of
the static spin susceptibility $\chi_{ij}(\star)$, where
$\star=({\vec q}\to 0,i\omega_n=0)$, reveal the complex ESOOP and
the breaking of the $C_{\infty v}$ symmetry\cite{Gorkov}. The
$\chi_{ij}(\star)$ is

\begin{equation}
\chi_{ij}(\star)=\mu_B^2 \lim_{{\vec q}\to 0}\int_{0}^{1/T} d\tau \,
\langle T_{\tau} \hat{m}_i({\vec q},\tau) \hat{m}_j(-{\vec q}, 0)\rangle
\label{susceptibility_1}
\end{equation}

\noindent where $\mu_B$ is the effective Bohr magneton, $\tau$ is
the Matsubara time, $T_{\tau}$ is the time ordering operator, $T$ is
the temperature, and $\hat{m}_i({\vec q},\tau)=\sum_{{\vec
k},\mu\,\nu}\hat{\alpha}^{\dagger}_{{\vec k}+{\vec q},\mu}(\tau)
\,[\sigma_i]^{\mu\,\nu}\,\hat{\alpha}_{{\vec k},\nu}(\tau)$ is the
magnetization operator in the lower excitonic branch. The Pauli
paramagnetic limit $\chi_P=2\mu_B^2 \nu_F$ is obtained expectedly
for the diagonal elements in the absence of SOC. Here $\nu_F$ is the
DOS at the Fermi level. We focus on the off diagonal terms in the
limit $T \to 0$, as those have the strongest signature of the
$C_{\infty v}$ breaking. Writing $\hat{\alpha}_{{\vec k},\mu}$ in
terms of $\eta_{{\vec k},\pm}$ in (\ref{spin_orbit_basis}), the
susceptibility can be written in terms of the Matsubara Green's
functions ${\cal G}_{{\vec k},\pm}(\tau)=-\langle T_{\tau}
\eta_{{\vec k},\pm}(\tau) \eta^\dagger_{{\vec k},\pm}(0)\rangle$.
For weak SOC we find

\begin{eqnarray}
&&\frac{\chi_{zx}(\star)+i\chi_{zy}(\star)}{\chi_P}\simeq -\frac{1}{4}\,
\frac{\partial}{\partial E_{\vec k}}
<iA+\Delta_1>_a \Big\vert_{E_{\vec k}=E_F}
\label{short_spin_suscept} \\
&&\frac{\chi_{xy}(\star)}{\chi_P} \simeq \frac{1}{6}\,\frac{\partial^2}{\partial E_{\vec k}^2}
\Im m\{<(iA+\Delta_1)^2>_a\}\Big\vert_{E_{\vec k}=E_F} \label{chi_xy}
\end{eqnarray}

\noindent 
where $<\dots>_a$ is the angular average and $\chi_P$ is the Pauli paramagnetic 
susceptibility. If the Fermi contour is isotropic, (\ref{short_spin_suscept}) and
(\ref{chi_xy}) both vanish. This occurs at low $n_x$ [i.e. (c) and (d) in
Figs \ref{robust_phase} and \ref{exc.energy}], where the phase of $\Delta_1$ is
coherent and $\vert \Delta_1 \vert$ is isotropic. On the other hand, at higher $n_x$
the Fermi contour is anisotropic [i.e. (a) and (b) in the same figures] and the
phase of $\Delta_1$ varies. Therefore, the effect in (\ref{short_spin_suscept})
and (\ref{chi_xy}) may be visible within the BCS limit at relatively high $n_x$.
Considering that the magnitude of
$\Delta_1$ is set by the e-h Coulomb interaction, we approximately have
$(\chi_{zx}+i\chi_{zy})/\chi_P \simeq V^{eh}_{{\vec q}_F}/E_F \le 0.1$ and
$\chi_{xy}/\chi_P \simeq (V^{eh}_{{\vec q}_F}/E_F)^2 \le 0.01$.

In conclusion, in the presence of excitonic background,
the interference between the electron and the hole SOCs renders
the e-h pairing unconventional by breaking the rotational symmetry of the ground
state. The resulting complex order parameter is affected by the exciton density. As
the density is increased, the magnitude smoothly changes from an isotropic BEC type
 to an anisotropic BCS type. On the other hand, its phase is globally coherent at
low densities, and gradually becomes nonuniform at increased densities.
The predicted strength is small but observable in the offdiagonal static spin
susceptibility; suggesting a new direction in the experimental observation of
the excitonic condensate.

We thank J. Schliemann for discussions. This research is supported by
T\"{U}B{\.I}TAK by the grants 105T110 and B.02.1.TBT.0.06.03.00/1120/2985.
The work on susceptibility was carried out at the
Marmaris Institute of Theoretical and Applied Physics (ITAP).

\newpage

\begin{thebibliography}{99}
\bibitem{footnote1} This approach is opposite to the case in Ref.[\cite{Gorkov}]
where SOC is stronger than the typical condensation energy.
\end{thebibliography}
\end{document}